\documentclass[a4paper,11pt]{article}
\pdfoutput=1 

\usepackage{jheppub} 

\usepackage[T1]{fontenc} 

\title{\boldmath Dilatonic Effect in Double Field Theory Cosmology}

\author[a]{Yang Liu}

\affiliation[a]{School of Physics and Astronomy, University of Nottingham, Nottingham NG7 2RD, UK}
\affiliation[b]{Nottingham Centre of Gravity, University of Nottingham, Nottingham NG7 2RD, UK}

\emailAdd{yang.liu@nottingham.ac.uk}

\usepackage{latexsym}

\abstract{In this article we discuss some aspects of double field theory cosmology with an emphasis on the role played by the dilaton. The cosmological solutions of double field theory equations of motion after coupling a shifted dilaton to them are investigated. The equations of motion for a constant shifted dilaton and a constant usaul dilaton in an FRW universe are obtained. The solutions of these equations are obtained in both the supergravity frame and in the winding frame. We also consider three possible dark energy candidates in a $4D$ universe using double field theory cosmology and find some basic conditions which the three dark energy candidates should satisfy. We consider the results for a more general potential of shifted dilaton as well.}

\begin{document} 
\maketitle
\flushbottom

\section{Introduction}
\label{sec:intro}
Despite its great success, General Relativity (GR) faces several shortcomings when applied to the universe.  In particular, the concepts of  dark matter and dark energy must be introduced in order to provide an explanation for  the large-scale dynamics of the universe. Furthermore, in order to solve the horizon and the flatness problems, new dynamics, such as inflation, involving additional degrees of freedom, which may play an important role near the strong-coupling regime at which GR breaks down, are required [1]. In order to make quantitative predictions concerning the early universe a consistent quantum gravity theory needs to be  developed.\\
At present, string theory appears to be the most promising candidate for quantum gravity theory. It is presently being used to study cosmology in the low energy effective supergravity (SUGRA) limit at weak coupling [2]. In GR, the only gravitational field is the spacetime metric $g_{\mu \nu}$. A Weyl transformation on the metric can be performed to convert it from the string frame to the Einstein frame.\\
The T-duality plays an important role in string theory [3]. It states that the physics of small compact spaces of radius $R$ is equivalent to the physics of large compact spaces of radius $1/R$ (in string units) [3]. For example, for strings on a torus of radius $R$, the symmetry implies that the spectrum of string states is unchanged if $R \rightarrow 1/R$ (in string units) and string momentum modes are interchanged with string winding modes. T-duality symmetry is assumed to be a fundamental symmetry of non-perturbative string theory [4].\\
The fields of double field theory (DFT) are $O(D,D)$ generalizations of spacetime fields [5,6,7,8,9]. The equivalence of spacetime momenta and winding numbers in the string spectra gives rise to a set of dual coordinates $\tilde{x}_i$, conjugated to winding numbers [10]. These dual coordinates are treated on the same footing as the usual coordinates $x_i$. Then spacetime dimension has changed from $D$ to $2D$. DFT is given (see, e.g., [11] for a review) by an action for a generalized metric in $2D$ space-time dimensions which is constructed from the metric, the antisymmetric tensor field and the dilaton (the “background”) of the massless sector of $D$ space-time dimensional string theory. In particular, after imposing a section condition the dynamical equations for the background reduce to those of supergravity. T-duality is already present in supergravity even in the absence of the $O(D,D)$ covariant structure introduced in DFT. However, T-duality is manifested as an $O(D,D)$ symmetry in the action of DFT [10]. The full set of coordinates in DFT can be denoted as $X^{M}(x^i, \tilde{x}^i)$, where $x^i (i= 1,2,...,D)$ is the usual spacetime coordinate, $\tilde{x}^i$ the dual coordinate and $M = 1,2,...,2D$ the $O(D,D)$ index. All of the spacetime component fields depend on both the usual and the dual coordinates, i.e., $\phi_I=\phi_I(x^i, \tilde{x}^i)$ [10]. \\
Cosmology in double field theory is a relatively new research field which has been investigated only recently and at present only a few papers are available on this topic.  DFT equations of motion for cosmology has been studied in ref.[10] but in the absence of sources of matter.  The authors demonstrate that the contraction of the conjugated space leads to both an inflation phase and a decelerated expansion of the ordinary space during different evolution stages [10]. Sources of matter have been included in the general equations derived in ref.[12].  Furthermore, ref.[3] and [13] considered the T-dual cosmological solutions. In ref.[14], Robert Brandenberger et al. made use of the T-duality symmetry of superstring theory and of the double geometry from double field theory and postulated that the cosmological singularities of a homogeneous and isotropic universe would disappear. Furthermore, Peng Wang et al. considered the effect of non-singular cosmology via $\alpha'$ corrections [15]. In addition, in ref.[16,17], the same authors considered the $O(D,D)$ duality string cosmology to all orders $\alpha'$. Further details are not listed here since these papers are not directly related to this article.\\ 
In the present article, we discuss some aspects of double field theory cosmology with an emphasis on the role played by the dilaton. The cosmological solutions from double field theory equations of motion after coupling a shifted dilaton to them are discussed in both a supergravity frame and a winding frame. In section 2 double field theory and dual cosmology are briefly reviewed. In section 3 the equations of motion (EQM) coupling a shifted dilaton in an FRW metric are obtained. In section 4 the solutions of the EQM for a constant  shifted dilaton in an FRW universe are considered. In section 5 the solutions for a constant usual diffeomorphic dilaton in an FRW universe are considered. In section 6 we consider three dark energy candidates in $4D$ DFT cosmology. In section 7 we analyze a more general potential of shifted dilaton. In section 8 conclusions and a discussion of future work are presented.\\
We adopt the following notations: capital letters $M$,$N$,... represent indices which encompass both regular and dual spacetime dimensions; small letters $i$, $j$,... denote indices which encompass regular $D = d +1$ spacetime dimensions. $a(t)$ is the cosmological scale factor, where $t$ is the physical time whilst the dual time is denoted by $\tilde{t}$. The $\omega = p/\rho $ is the equation of state, where $p$ and $\rho$ are pressure and energy density, respectively.

\section{Short review of double field theory and dual cosmology}
\subsection{Double field theory}
Current research has primarily been focused on the massless sector of closed string spectra [10]. The DFT action includes a $D$ dimensional metric $g_{ij}$, the anti-symmetric Kalb-Ramond field $b_{ij}$ and the dilaton field $\phi$. The action rewrites these fields in an $O(D,D)$ covariant way, where $D$ is the dimensionality of space-time. If the fields were only dependent on the usual coordinates, the DFT action could be reduced to a supergravity action [3]. The DFT action is given by [3]
\begin{equation}\label{eq:2.1}
S = \int dx d\tilde{x} e^{-2d} \mathcal{R},
\end{equation}   
where $d$ contains the usual dilaton $\phi$ and the determinant of the metric $g$, i.e.
\begin{equation}\label{eq:2.2}
e^{-2d} = \sqrt{-g} e^{-2\phi},
\end{equation} 
and [10]
\begin{equation}\label{eq:2.3}
\mathcal{R}= \frac{1}{8} \mathcal{H}^{MN} \partial_{M} \mathcal{H}^{KL} \partial_{N} \mathcal{H}_{KL} - \frac{1}{2} \mathcal{H}^{MN} \partial_{N} \mathcal{H}^{KL} \partial_{L} \mathcal{H}_{MK} - \partial_{M} d \partial_{N} \mathcal{H}^{MN} + 4 \mathcal{H}^{MN} \partial_{M} d  \partial_{N} d,
\end{equation}
where the generalized metric, $\mathcal{H}_{MN}$, is defined as
\begin{equation}\label{eq:2.4}
\mathcal{H}_{MN}
=
\begin{bmatrix}
g^{ij} & -g^{ik} b_{kj}\\
b_{ik}g^{kj} & g_{ij}- b_{ik} g^{kl} b_{lj}
\end{bmatrix}.
\end{equation}
The level matching condition in closed string theory imposes the “weak constraint” $\partial \tilde{\partial} \phi (x, \tilde{x}) = 0$ for any field $\phi (x, \tilde{x})$. To ensure that the action is locally equivalent to the low energy effective string action, the following so-called “strong constraint” is required: $\partial \tilde{\partial} = 0$ as an operator equation, acting on any products of the fields [10]. 

\subsection{Dual cosmology}
In this article, we will set the Kalb-Ramond field $b_{ij} =0$ and adopt the following FRW like metric:
\begin{equation}\label{eq:2.5}
dS^2 = -dt^2 -d\tilde{t}^2 + a^2(t,\tilde{t}) dx^2 + \tilde{a}^2 (t,\tilde{t}) d\tilde{x}^2,
\end{equation} 
where $\tilde{a} = a^{-1}$ [13]. \\
The vacuum equations of motion for DFT in the presence of a dual time associated with the winding sector are given by [10]
\begin{equation}\label{eq:2.6}
4 d^{''} - 4d^{'2} -(D-1)\tilde{H}^2 + 4 \ddot{d} - 4 \dot{d}^2 - (D-1)H^2 =0,
\end{equation} 
\begin{equation}\label{eq:2.7}
(D-1)\tilde{H}^2 -2d^{''} - (D-1)H^2 + 2\ddot{d} = 0,
\end{equation}
\begin{equation}\label{eq:2.8}
\tilde{H}^{'} -2\tilde{H}d^{'} + \dot{H} - 2H\dot{d} =0,
\end{equation}
where a prime denotes a derivative with respect to $\tilde{t}$, and a dot denotes a derivative with respect to $t$. Noting that $2d = 2\phi - (D-1)\ln a$ defines the shifted dilaton and $\tilde{H} = a'/a$, by variation of action $(2.1)$ with respect to $d$, we can obtain the equation $(2.6)$ for a shifted dilaton and by variation of action $(2.1)$ with respect to the generalized metric $\mathcal{H}_{MN}$, we can obtain eqs.$(2.7)$ and $(2.8)$ for a graviton. \\
In ref.[12], the authors proposed the following cosmological equations in the presence of matter by the following prescription:
\begin{equation}\label{eq:2.9}
4 d^{''} - 4d^{'2} -(D-1)\tilde{H}^2 + 4 \ddot{d} - 4 \dot{d}^2 - (D-1)H^2 =0,
\end{equation} 
\begin{equation}\label{eq:2.10}
(D-1)\tilde{H}^2 -2d^{''} - (D-1)H^2 + 2\ddot{d} = \frac{1}{2} e^{2d} E(t, \tilde{t}),
\end{equation}
\begin{equation}\label{eq:2.11}
\tilde{H}^{'} -2\tilde{H}d^{'} + \dot{H} - 2H\dot{d} =\frac{1}{2} e^{2d} P(t, \tilde{t}),
\end{equation}
where $E$ and $P$ are the energy and pressure associated with the matter sector. \\
In the following, we will consider the stabilized dilaton $\phi = \phi_0$. In this case we have $2\dot{d}= -(D-1)H$ and $2d^{'} = (D-1)\tilde{H}$, and eqs.$(2.9)-(2.11)$ become
\begin{equation}\label{eq:2.12}
2(\tilde{H}^{'} - \dot{H}) - D(\tilde{H}^{2} + H^2) =0,
\end{equation} 
\begin{equation}\label{eq:2.13}
(\tilde{H}^2-H^2)-(\tilde{H}^{'}+\dot{H})= \frac{1}{2(D-1)} G \rho(t,\tilde{t}),
\end{equation}
\begin{equation}\label{eq:2.14}
(\tilde{H}^{'} + \dot{H}) + (D-1) ( H^2-\tilde{H}^{2} ) = \frac{1}{2} G p(t,\tilde{t}),
\end{equation}
where $G$ depends on $\phi =\phi_0$ which is the fixed value of a dilaton [3]. \\
The mass spectrum of a closed string in one dimensional space, compactified on a circle, is given by [3]
\begin{equation}\label{eq:2.15}
 M^2 = (N + \tilde{N} -2) + p^2 \frac{l^2_s}{R^2} + \omega^2 \frac{l^2_s}{\tilde{R}^2}, 
\end{equation}
where $N$, $\tilde{N}$ are the oscillaratory modes of the string, $p$ denotes the momentum modes which associated with the center of mass, and $\omega$ denotes the winding modes which represent the number of times the string has wrapped itself around the compact dimension in a topologically nontrivial way [3]. Taking the limit of large $R$ all physical quantities depend solely on $t$ which identifies the so-called “supergravity frame”. Taking the limit of small $R$, all physical quantities depend solely on $\tilde{t}$, which identifies the so-called “winding frame” [3]. 

\section{The equations of motion coupling a shifted dilaton in an FRW like universe}
Following ref.[18], we will add a dilaton potential $V(d)$ into the action $(2.1)$, so that
\begin{equation}\label{eq:3.1}
S = \int dx d\tilde{x} e^{-2d} \{\mathcal{R} - 2V(d) \} + S_m, 
\end{equation} 
where $S_m$ is the action of the matter sector. The manner of introducing matter is described in Appendix. The reason for considering the effect of a dilaton in double  field theory cosmology  is that, in standard string cosmology, a dilaton is often regarded as  quintessence (such as ref.[18]), which can be used to explain the origin of inflation and the accelerating expansion of the universe. In this article, we will consider whether a shifted dilaton can be used to account for the real physical phenomena.\\
In order to avoid the singularity, we introduce a dilaton potential
\begin{equation}\label{eq:3.2}
V(d) = V_0 e^{8d}, 
\end{equation} 
where $V_0 > 0$ [10]. It is known that non-singular solutions can be obtained at low curvatures in the presence of an approriate non-local effective potential [19]. This non-local potential represents the back reactions of higher loop corrections [19]. $V_0$ includes a proper volume which makes the dilaton potential a scalar under generalized diffeomorphisms. This potential guarantees the $O(D,D)$ symmetry [10]. In fact, $V_0$ cannot be negative. We should point out two points. Firstly, $V_0$ is a volume [14,19] which cannot be zero. Moreover, in ref.[19], the Hubble parameter $t^{-1}_0 = e^{\phi_0} \sqrt{V_0}$, where $\phi_0$ is an integration constant. If $V_0$ is negative, then the result is unphysical. However, the prefactor of potential can be negative, i.e., the potential can be $V(d) =-V_0 e^{8d}$, where $V_0 >0$. But in sections 3, 4, 5 and 6, we only consider potential eq.$(3.2)$, which can be regarded as a particular example. In section 7, we will consider a more general potential $V(d) =A e^{nd}$, where $A$ and $n$ can be arbitrary real numbers. \\
By variation of the shifted dilaton $d$ of action $(3.2)$, we can obtain the equation of motion of the shifted dilaton
\begin{equation}\label{eq:3.3}
4 d^{''} - 4d^{'2} -(D-1)\tilde{H}^2 + 4 \ddot{d} - 4 \dot{d}^2 - (D-1)H^2 - 12 V_0 e^{6d}=0.
\end{equation}      
By variation of the generalized  metric $\mathcal{H}^{MN}$ of action $(3.2)$, the equations of motion of graviton are the same as eq.$(2.10)$ and $(2.11)$ so that the complete list of the equations of motion are:
\begin{equation}\label{eq:3.4}
4 d^{''} - 4d^{'2} -(D-1)\tilde{H}^2 + 4 \ddot{d} - 4 \dot{d}^2 - (D-1)H^2 - 12 V_0 e^{6d}=0,
\end{equation}
\begin{equation}\label{eq:3.5}
(D-1)\tilde{H}^2 -2d^{''} - (D-1)H^2 + 2\ddot{d} = \frac{1}{2} e^{2d} E(t, \tilde{t}),
\end{equation}
\begin{equation}\label{eq:3.6}
\tilde{H}^{'} -2\tilde{H}d^{'} + \dot{H} - 2H\dot{d} =\frac{1}{2} e^{2d} P(t, \tilde{t}).
\end{equation} 
In a supergravity frame, all physical quantities depend solely on physical time $t$:
\begin{equation}\label{eq:3.7}
 4 \ddot{d} - 4 \dot{d}^2 - (D-1)H^2 - 12 V_0 e^{6d}=0,
\end{equation}
\begin{equation}\label{eq:3.8}
 - (D-1)H^2 + 2\ddot{d} = \frac{1}{2} e^{2d} E(t),
\end{equation}
\begin{equation}\label{eq:3.9}
 \dot{H} - 2H\dot{d} =\frac{1}{2} e^{2d} P(t).
\end{equation}
In a winding frame, all physical quantities depend solely on dual time $\tilde{t}$:
\begin{equation}\label{eq:3.10}
4 d^{''} - 4d^{'2} -(D-1)\tilde{H}^2  - 12 V_0 e^{6d}=0,
\end{equation}
\begin{equation}\label{eq:3.11}
(D-1)\tilde{H}^2 -2d^{''} = \frac{1}{2} e^{2d} E(\tilde{t}),
\end{equation}
\begin{equation}\label{eq:3.12}
\tilde{H}^{'} -2\tilde{H}d^{'}  =\frac{1}{2} e^{2d} P(\tilde{t}).
\end{equation}
In the following sections, we will consider the solutions of the above equations for constant shifted dilaton $d_0$ and a constant usual dilaton $\phi_0$ in both a supergravity frame as well as in a winding frame. 

\section{Solutions of the equations of motion for a constant shifted dilaton in an FRW like universe}
In this section, we will obtain the solutions of the equations of motion for a constant dilaton.\\
For a constant shifted dilaton $d$ in a supergravity frame, i.e., $\dot{d} = 0$, $d=d_0$, then the equations of motion become
\begin{equation}\label{eq:4.1}
 - (D-1)H^2 - 12 V_0 e^{6d_0}=0,
\end{equation}
\begin{equation}\label{eq:4.2}
- (D-1)H^2 = \frac{1}{2} e^{2d_0} E(t),
\end{equation}
\begin{equation}\label{eq:4.3}
\dot{H} =\frac{1}{2} e^{2d_0} P(t).
\end{equation}
From eq.$(4.1)$, we obtain
\begin{equation}\label{eq:4.4}
H = \sqrt{\frac{-12V_0}{D-1}} e^{3d_0},
\end{equation}
and when $V_0>0$, $H$ is a complex number, therefore there is no physical solution.\\
In a similar manner to that presented above, we can obtain the corresponding solutions in a winding frame:
\begin{equation}\label{eq:4.5}
\tilde{H} = \sqrt{\frac{-12V_0}{D-1}} e^{3d_0},
\end{equation}
which is also a complex number as in the supergravity frame, therefore the solution is unphysical.

\section{Solutions of the equations of motion for a constant usual dilaton in an FRW like universe}
In this section, we briefly consider the T-dual solutions in both a supergravity frame and a winding frame firstly. Then we will consider the solutions of the equations of motion for a constant usual dilaton $\phi$ in both a supergravity frame and in a winding frame.\\
For constant $\phi = \phi_0$, considering T-dual solutions, then we have [13]:
\begin{equation}\label{eq:5.1}
2\dot{d} = \alpha (D-1) H,
\end{equation}
\begin{equation}\label{eq:5.2}
2d^{'} = \tilde{\alpha} (D-1) \tilde{H},
\end{equation}
where $\alpha$ and $\tilde{\alpha}$ are both constants. In particular, for $(\alpha, \tilde{\alpha}) = (-1, 1)$ we have a constant usual dilaton in the supergravity frame and a non-constant usual dilaton in the winding frame; for $(\alpha, \tilde{\alpha}) = (1, -1)$ we have a constant usual dilaton in the winding frame and nonconstant usual dilaton in the supergravity frame. The case of $(\alpha, \tilde{\alpha}) = (-1, -1)$ corresponds to having a constant usual dilaton in both frames [13].\\ 

\subsection{Solutions in a supergravity frame}
We obtain respective sets of equations for the supergravity frame and the winding frame. In the supergravity frame, the equations of motion are:
\begin{equation}\label{eq:5.3}
2\alpha (D-1) \dot{H} - H^2 (D-1) [\alpha^2 (D-1) +1] - 12 V_0 e^{6d} = 0,
\end{equation}
\begin{equation}\label{eq:5.4}
-(D-1)H^2 + \alpha (D-1) \dot{H} = \frac{1}{2} e^{2d} E(t),
\end{equation}
\begin{equation}\label{eq:5.5}
\dot{H} - \alpha (D-1) H^2 = \frac{1}{2} e^{2d} P(t),
\end{equation}
therefore
\begin{equation}\label{eq:5.6}
H^2 = \frac{12V_0 e^{6d} [\omega - \frac{1}{\alpha (D-1)}]}{[1- \alpha^2(D-1)] [\frac{1}{\alpha} + \omega(D-1)]}.
\end{equation}
Applying the section conditions, then for stabilized $\phi = \phi_0$, $2\dot{d} = -(D-1)H$ and $2\ddot{d}= -(D-1)\dot{H}$, then the eq.$(3.7)$ becomes
\begin{equation}\label{eq:5.7}
2(D-1)\dot{H} + (D-1)^2 H^2 + (D-1)H^2 + 12V_0 e^{6d} =0,
\end{equation}
namely,
\begin{equation}\label{eq:5.8}
\dot{H} = - \frac{1}{2(D-1)} \{H^2[(D-1)^2 + (D-1)] + 12V_0 e^{6d} \}.
\end{equation} 
Inserting eq.$(5.8)$ into eq.$(3.8)$ and $(3.9)$, we have
\begin{equation}\label{eq:5.9}
\frac{1}{2} H^2 (D^2 - 3D +2)  + 6V_0 e^{6d} =\frac{1}{2} e^{2d} E(t),
\end{equation} 
\begin{equation}\label{eq:5.10}
\frac{1}{2} H^2 (D-2)  - \frac{6V_0}{(D-1)} e^{6d} =\frac{1}{2} e^{2d} P(t).
\end{equation} 
Combining eq.$(5.9)$ and $(5.10)$, we obtain that
\begin{equation}\label{eq:5.11}
 H^2 = \frac{(D-1)\omega + 1}{2(D-1)(D-2)} e^{2d} E(t) = \frac{(D-1)\omega + 1}{2(D-1)(D-2)} e^{2\phi_0} a^{-(D-1)} E(t),
\end{equation} 
\begin{equation}\label{eq:5.12}
E(t)=\frac{24V_0}{1-\omega(D-1)} e^{4\phi_0} a^{-2(D-1)},
\end{equation}
where $\omega$ is the equation of state. From eq.$(5.11)$, it is apparent $D$ cannot be 1 and 2 unless $\omega= - \frac{1}{D-1}$. \\
When $\omega = \frac{1}{D-1}$, $V(d) =0$, and
\begin{equation}\label{eq:5.13}
H^2 = \frac{1}{(D-1)(D-2)} e^{2\phi_0} a^{-(D-1)} E(t),
\end{equation}
which corresponds to the case with no dilaton potential [3].\\
In particular, for $D=4$, if $\omega = \frac{1}{D-1} = \frac{1}{3}$, where radiation dominates the universe, then
\begin{equation}\label{eq:5.14}
H^2 = \frac{1}{6} e^{2\phi_0} a^{-3} E(t).
\end{equation}
Moreover, if we assume $\omega$ is a constant, considering eq.$(5.11)$ and $(5.12)$, then we obtain
\begin{equation}\label{eq:5.15}
a = \left(\frac{9}{4}(D-1)^2\right)^{\frac{1}{3(D-1)}} t^{\frac{2}{3(D-1)}},
\end{equation}
where we have set integration constant to be zero. Then
\begin{equation}\label{eq:5.16}
\dot{a} = \left(\frac{9}{4}(D-1)^2\right)^{\frac{1}{3(D-1)}} \frac{2}{3(D-1)} t^{\frac{5-3D}{3(D-1)}},
\end{equation}
\begin{equation}\label{eq:5.17}
H=\dot{a}/a = \frac{2}{3(D-1)} t^{-1},
\end{equation}
\begin{equation}\label{eq:5.18}
\ddot{a} = \left(\frac{9}{4}(D-1)^2\right)^{\frac{1}{3(D-1)}} \frac{10-6D}{9(D-1)^2} t^{\frac{8-6D}{3(D-1)}}. 
\end{equation}
In order to account for the accelerating expansion of the universe, $\ddot{a}$ must be larger than $0$. However, this is impossible for $D \neq 1$.\\

\subsection{Solutions in a winding frame}
Similarly, in the winding frame, the equations of motion are:
\begin{equation}\label{eq:5.19}
2 \tilde{\alpha} (D-1) \tilde{H}^{'} - \tilde{H}^2 (D-1) [\tilde{\alpha}^2 (D-1) +1] - 12 V_0 e^{6d} = 0,
\end{equation}
\begin{equation}\label{eq:5.20}
(D-1)\tilde{H}^2 - \tilde{\alpha} (D-1) \tilde{H}^{'} = \frac{1}{2} e^{2d} E(\tilde{t}),
\end{equation}
\begin{equation}\label{eq:5.21}
\tilde{H}^{'} - \tilde{\alpha} (D-1) \tilde{H}^2 = \frac{1}{2} e^{2d} P(\tilde{t}),
\end{equation}
therefore
\begin{equation}\label{eq:5.22}
\tilde{H}^2 = \frac{12V_0 e^{6d} [\omega + \frac{1}{\tilde{\alpha} (D-1)}]}{[1- \tilde{\alpha}^2(D-1)] [\omega(D-1) - \frac{1}{\tilde{\alpha}}]}.
\end{equation}
For stabilized $\phi = \phi_0$, $2d =2\phi_0 - (D-1)\ln a$ and $2d^{'} = (D-1)\tilde{H}$, we obtain
\begin{equation}\label{eq:5.23}
\tilde{H}^2 = \frac{[(D-1)\omega + 1]}{2(2-D)(D-1)} e^{2d} E(\tilde{t}) = \frac{[(D-1)\omega +1]}{2(2-D)(D-1)} e^{2\phi_0} a^{-(D-1)} E(\tilde{t}),
\end{equation}
\begin{equation}\label{eq:5.24}
E(\tilde{t}) = \frac{24V_0}{1-\omega(D-1)} e^{4\phi_0} a^{-2(D-1)},
\end{equation}
where $\omega$ is the equation of state. From eq.$(5.23)$, it is apparent $D$ cannot be 1 and 2 unless $\omega = -\frac{1}{D-1}$. \\
When $\omega =  \frac{1}{D-1}$, $V(d)=0$, then
\begin{equation}\label{eq:5.25}
\tilde{H}^2 = -\frac{1}{(D-2)(D-1)} e^{2d} E(\tilde{t}) = -\frac{1}{(D-2)(D-1)} e^{2\phi_0} a^{-(D-1)} E(\tilde{t}),
\end{equation}
which is the case without a dilaton potential [3].\\
In particular, for $D=4$, if $\omega = 1/3$, then 
\begin{equation}\label{eq:5.26}
\tilde{H}^2 = -\frac{1}{6} e^{2d} E(\tilde{t}) = -\frac{1}{6} e^{2\phi_0} a^{-3} E(\tilde{t}).
\end{equation} 
Moreover, if we assume $\omega$ is a constant, considering eq.$(5.24)$ and $(5.25)$, then we obtain
\begin{equation}\label{eq:5.27}
a = \left(\frac{9}{4}(D-1)^2\right)^{\frac{1}{3(D-1)}} \tilde{t}^{\frac{2}{3(D-1)}},
\end{equation}
where we have set integration constant to be zero. Then
\begin{equation}\label{eq:5.28}
a' = \left(\frac{9}{4}(D-1)^2\right)^{\frac{1}{3(D-1)}} \frac{2}{3(D-1)} \tilde{t}^{\frac{5-3D}{3(D-1)}},
\end{equation}
\begin{equation}\label{eq:5.29}
H=a'/a = \frac{2}{3(D-1)} \tilde{t}^{-1},
\end{equation}
\begin{equation}\label{eq:5.30}
a'' = \left(\frac{9}{4}(D-1)^2\right)^{\frac{1}{3(D-1)}} \frac{10-6D}{9(D-1)^2} \tilde{t}^{\frac{8-6D}{3(D-1)}}. 
\end{equation}

\section{Candidates for dark energy}
Since $V_0 > 0$, in order to ensure that $a$ is real, according to $(5.16)$, we must require
\begin{equation}\label{eq:6.1}
\frac{\frac{1}{D-1} + \omega}{(D-2)[1-\omega (D-1)]} > 0.
\end{equation}
From eq.$(6.1)$, we obtain two “critical” dimensions. One is $D=2$, and the other one is $D= 1 + \frac{1}{\omega}$. Therefore, for “critical” dimensions, $\omega = \frac{1}{n}$, where $n$ denotes positive integers.\\
From eq.$(6.1)$, we can obtain the constraints for $\omega$ to be
\begin{equation}\label{eq:6.2}
- \frac{1}{D-1} < \omega < \frac{1}{D-1}.
\end{equation}
Therefore, for the real universe, i.e., $D=4$ the constraints on $\omega$ are,  $-1/3 < \omega < 1/3$. Based on ref.[22], we know that for any general spacetime dimension $D$, tensile matter has the equation of state
\begin{equation}\label{eq:6.3}
p = \omega \rho.
\end{equation}
In order to explain the origin of the accelerating expansion of the universe [22]
\begin{equation}\label{eq:6.4}
-1 < \omega < -(\frac{D-3}{D-1}),
\end{equation}
namely, $-1 < \omega < -1/3$. We therefore conclude that this model with a constant usual $\phi = \phi_0$ cannot be used as an explanation for the accelerating expansion of the universe.\\
We will now consider three possible dark energy candidates in $D = 4$ double field theory cosmology.\\
Firstly, for a holographic dark energy scenario $\rho_D = 3c^2 m^2_p /L^2$, where $c$ is a dimensionless parameter (usually we take $c>0$), $m^2_p$ is Planck mass and $L = ar(t) =R_h$, where $R_h$ is the radial size of the horizon [23]. If an interaction between matter and holographic dark energy exists, i.e.
\begin{equation}\label{eq:6.5}
\dot{\rho}_D + 3H \rho_D (1+\omega_D) = -Q,
\end{equation}
\begin{equation}\label{eq:6.6}
\dot{\rho}_m + 3H \rho_m = Q,
\end{equation}
where $Q = 3b^2H(\rho_m + \rho_D)$, $b^2$ is a coupling parameter and $\rho_m$ and $\rho_D$ are the energy densities of matter and dark energy respectively [24] then
\begin{equation}\label{eq:6.7}
\omega_D = -\frac{1}{3} - \frac{2\sqrt{\Omega_D}}{3c} - \frac{b^2}{\Omega_D},
\end{equation}
where $\omega_D$ is the equation of state of dark energy, $c$ is a dimensionless parameter and $\Omega_D$ is the fractional energy density which is defined as:
\begin{equation}\label{eq:6.8}
\Omega_D = \frac{\rho_D}{\rho_{cr}},
\end{equation}
where $\rho_{cr} = 3 m^2_p H^2$. Therefore $\omega_D < -\frac{1}{3}$ when $c>0$. Considering eq.$(6.2)$, so that in view of eq.$(6.2)$ a holographic dark energy scenario cannot exist in $D=4$ DFT cosmology.\\
Secondly, for a ghost dark energy scenario, $\rho_D =\alpha H$, where $\alpha$ denotes a constant of order $\Lambda^3_{QCD}$ and QCD represents the QCD mass scale and $H$ is the Hubble constant. Taking the same interaction $Q$ between matter and dark energy as in eq.$(6.5)$ and $(6.6)$, we have [25]
\begin{equation}\label{eq:6.9}
\omega_D = -\frac{1}{2-\Omega_D} (1+ \frac{2b^2}{\Omega_D}),
\end{equation}
therefore
\begin{equation}\label{eq:6.10}
\frac{\frac{1}{3} + \omega_D}{1-3\omega_D} = -\frac{1+\Omega_D + \frac{6b^2}{\Omega_D}}{3(5-\Omega_D+ \frac{6b^2}{\Omega_D})} < 0,
\end{equation}
since $0 < \Omega_D < 1$. According to eq.$(5.16)$, $A_D$ is a complex number so that in view of eq.$(6.2)$, the ghost dark energy scenario cannot exist in $D=4$ DFT cosmology.\\
Thirdly, for a Tsallis holographic dark energy scenario $\rho_D = B H^{-2\delta + 4}$, where $B$ is an unknown parameter and $\delta$ represents the non-additivity parameter [26]. Taking the same interaction $Q$ between matter and dark energy as in eq.$(6.5)$ and $(6.6)$, we have: [27]
\begin{equation}\label{eq:6.11}
\omega_D = \frac{\delta-1 + b^2/ \Omega_D}{(2-\delta)\Omega_D - 1}.
\end{equation}
Considering eq.$(6.1)$, if Tsallis holographic energy can exist in $D=4$ double field theory cosmology of the universe, then the parameters must satisfy the inequalities:
\begin{equation}\label{eq:6.12}
(2-\delta) \Omega^2_D + (3\delta -4) \Omega_D + 3b^2 > 0,
\end{equation}
\begin{equation}\label{eq:6.13}
(2-\delta) \Omega^2_D + (-3\delta +2) \Omega_D + 3b^2 > 0,
\end{equation}
or
\begin{equation}\label{eq:6.14}
(2-\delta) \Omega^2_D + (3\delta -4) \Omega_D + 3b^2 < 0,
\end{equation}
\begin{equation}\label{eq:6.15}
(2-\delta) \Omega^2_D + (-3\delta +2) \Omega_D + 3b^2 < 0.
\end{equation}

\section{Analysis of a more general potential}
In this section, we will analyze a more general potential $V(d)=A e^{nd}$, where $A$ and $n$ are arbitrary real numbers. We will see how the results in sections 3, 4 and 5 change. \\ 

\subsection{Case 1: A>0}
When $A>0$, the potential can be written as $V(d)=V_0 e^{nd}$, where $V_0>0$. \\
In a supergravity frame, the equations of motion are:
\begin{equation}\label{eq:7.1}
4 \ddot{d} - 4 \dot{d}^2 - (D-1)H^2 - 2(n-2) V_0 e^{(n-2)d}=0,
\end{equation}
\begin{equation}\label{eq:7.2}
- (D-1)H^2 + 2\ddot{d} = \frac{1}{2} e^{2d} E(t),
\end{equation}
\begin{equation}\label{eq:7.3}
\dot{H} - 2H\dot{d} =\frac{1}{2} e^{2d} P(t).
\end{equation}
In a winding frame, the equations of motion are:
\begin{equation}\label{eq:7.4}
4 d^{''} - 4d^{'2} -(D-1)\tilde{H}^2  - 2(n-2) V_0 e^{(n-2)d}=0,
\end{equation}
\begin{equation}\label{eq:7.5}
(D-1)\tilde{H}^2 -2d^{''} = \frac{1}{2} e^{2d} E(\tilde{t}),
\end{equation}
\begin{equation}\label{eq:7.6}
\tilde{H}^{'} -2\tilde{H}d^{'}  =\frac{1}{2} e^{2d} P(\tilde{t}).
\end{equation}

\subsubsection{Solutions for a constant shifted dilaton}
In a supergravity frame, a constant shifted dilaton means $\dot{d}=0$ and $d=d_0$. Then according to eq.$(7.1)$-$(7.3)$, we obtain 3 cases:\\
if $n<2$, then 
\begin{equation}\label{eq:7.7}
H=\sqrt{\frac{-2(n-2)V_0 e^{(n-2)d_0}}{D-1}}, \qquad E= 4(n-2)V_0 e^{(n-4)d_0}, \qquad P=0,
\end{equation}
which are physical solutions;\\
if $n=2$, then $H=0$, $E=0$ and $P=0$;\\
if $n>2$, then the solutions are unphysical.\\
In a winding frame, a constant shifted dilaton means $d'=0$ and $d=d_0$. Then according to eq.$(7.4)$-$(7.6)$, we obtain 3 cases as well:\\
if $n<2$, then 
\begin{equation}\label{eq:7.8}
\tilde{H}=\sqrt{\frac{-2(n-2)V_0 e^{(n-2)d_0}}{D-1}}, \qquad E= -4(n-2)V_0 e^{(n-4)d_0}, \qquad P=0,
\end{equation}
which are physical solutions;\\
if $n=2$, then $H=0$, $E=0$ and $P=0$;\\
if $n>2$, then the solutions are unphysical.\\

\subsubsection{Solutions for a constant usual dilaton}
In a supergravity frame, a constant usual dilaton means $\phi =\phi_0$, $2\dot{d} =-(D-1)H$ and $2\ddot{d}=-(D-1)\dot{H}$. Then according to eq.$(7.1)$ to $(7.3)$, we have
\begin{equation}\label{eq:7.9}
H^2 = \frac{1+\omega(D-1)}{2(D-1)(D-2)} e^{2d} E(t),
\end{equation}
and 
\begin{equation}\label{eq:7.10}
E(t) = \frac{4(n-2)V_0 e^{(n-4)d}}{1-\omega(D-1)}, \qquad P(t)=\omega E(t).
\end{equation}
In a winding frame, a constant usual dilaton means $\phi =\phi_0$, $2d' =(D-1)\tilde{H}$ and $2d''=(D-1)\tilde{H}'$. Then according to eq.$(7.4)$ to $(7.6)$, we have
\begin{equation}\label{eq:7.11}
\tilde{H}^2 = \frac{1+\omega(D-1)}{2(D-1)(D-2)} e^{2d} E(\tilde{t}),
\end{equation}
and 
\begin{equation}\label{eq:7.12}
E(\tilde{t}) = \frac{4(n-2)V_0 e^{(n-4)d}}{\omega(D-1)-1}, \qquad P(\tilde{t})=\omega E(\tilde{t}).
\end{equation}

\subsection{Case 2: A<0}
When $A<0$, we can rewrite the potential as $V(d)=-V_0e^{nd}$, where $V_0>0$.\\ 
In a supergravity frame, the equations of motion are:
\begin{equation}\label{eq:7.13}
4 \ddot{d} - 4 \dot{d}^2 - (D-1)H^2 + 2(n-2) V_0 e^{(n-2)d}=0,
\end{equation}
\begin{equation}\label{eq:7.14}
- (D-1)H^2 + 2\ddot{d} = \frac{1}{2} e^{2d} E(t),
\end{equation}
\begin{equation}\label{eq:7.15}
\dot{H} - 2H\dot{d} =\frac{1}{2} e^{2d} P(t).
\end{equation}
In a winding frame, the equations of motion are:
\begin{equation}\label{eq:7.16}
4 d^{''} - 4d^{'2} -(D-1)\tilde{H}^2  + 2(n-2) V_0 e^{(n-2)d}=0,
\end{equation}
\begin{equation}\label{eq:7.17}
(D-1)\tilde{H}^2 -2d^{''} = \frac{1}{2} e^{2d} E(\tilde{t}),
\end{equation}
\begin{equation}\label{eq:7.18}
\tilde{H}^{'} -2\tilde{H}d^{'}  =\frac{1}{2} e^{2d} P(\tilde{t}).
\end{equation}

\subsubsection{Solutions for a constant shifted dilaton}
In a supergravity frame, a constant shifted dilaton means $\dot{d}=0$ and $d=d_0$. Then according to eq.$(7.13)$-$(7.15)$, we obtain 3 cases:\\
if $n>2$, then 
\begin{equation}\label{eq:7.19}
H=\sqrt{\frac{2(n-2)V_0 e^{(n-2)d_0}}{D-1}}, \qquad E= -4(n-2)V_0 e^{(n-4)d_0}, \qquad P=0,
\end{equation}
which are physical solutions;\\
if $n=2$, then $H=0$, $E=0$ and $P=0$;\\
if $n<2$, then the solutions are unphysical.\\
In a winding frame, a constant shifted dilaton means $d'=0$ and $d=d_0$. Then according to eq.$(7.16)$-$(7.18)$, we obtain 3 cases as well:\\
if $n>2$, then 
\begin{equation}\label{eq:7.20}
\tilde{H}=\sqrt{\frac{2(n-2)V_0 e^{(n-2)d_0}}{D-1}}, \qquad E= 4(n-2)V_0 e^{(n-4)d_0}, \qquad P=0,
\end{equation}
which are physical solutions;\\
if $n=2$, then $H=0$, $E=0$ and $P=0$;\\
if $n<2$, then the solutions are unphysical.\\

\subsubsection{Solutions for a constant usual dilaton}
In a supergravity frame, a constant usual dilaton means $\phi =\phi_0$, $2\dot{d} =-(D-1)H$ and $2\ddot{d}=-(D-1)\dot{H}$. Then according to eq.$(7.13)$ to $(7.15)$, we have
\begin{equation}\label{eq:7.21}
H^2 = \frac{1+\omega(D-1)}{2(D-1)(D-2)} e^{2d} E(t),
\end{equation}
and 
\begin{equation}\label{eq:7.22}
E(t) = \frac{-4(n-2)V_0 e^{(n-4)d}}{1-\omega(D-1)}, \qquad P(t)=\omega E(t).
\end{equation}
In a winding frame, a constant usual dilaton means $\phi =\phi_0$, $2d' =(D-1)\tilde{H}$ and $2d''=(D-1)\tilde{H}'$. Then according to eq.$(7.16)$ to $(7.18)$, we have
\begin{equation}\label{eq:7.23}
\tilde{H}^2 = -\frac{1+\omega(D-1)}{2(D-1)(D-2)} e^{2d} E(\tilde{t}),
\end{equation}
and 
\begin{equation}\label{eq:7.24}
E(\tilde{t}) = \frac{-4(n-2)V_0 e^{(n-4)d}}{\omega(D-1)-1}, \qquad P(\tilde{t})=\omega E(\tilde{t}).
\end{equation}

\section{Conclusion and discussion}
The T-duality plays an important role in string theory [3]. Due to the equivalence of spacetime momenta and winding numbers in the string spectra, a set of dual coordinates $\tilde{x}^i$, which is conjugated to winding numbers [10] has been introduced in double field theory (DFT). These dual coordinates are treated on the same footing as the usual coordinates $x^i$. Therefore, the dimensionality of spacetime is from $D$ to $2D$ naturally. \\
Cosmology in double field theory is a relatively new research field which has received attention only in the last few recent years. At present there are relatively few papers dealing with this topic. The equations of motion for double field theory cosmology have been studied in ref.[10] in the situation where sources of matter are absent. Furthermore, the situation where sources of matter are present has have been discussed in the general equations derived in ref.[12]. T-dual cosmological solutions have also been dealt with in refs.[3], [13] and [29]. \\
In this article, we have discussed some aspects of double field theory cosmology with an emphasis on the role played  by  the dilaton and have investigated the cosmological solutions from double field theory equations of motion after coupling a shifted dilaton to them. In fact, many researches have been done to study the properties of dilaton potential in classical supergravity and double field theory to higher loop orders, such as refs.[30,31]. However, the dilatonic effect in DFT cosmology has not been discussed. In this acticle, we only consider the effect to lowest order. The equations of motion coupling a shifted dilaton in an FRW universe have been obtained in section 3. In section 4, we obtained the solutions of the equations of motion for a constant shifted dilaton in both a supergravity frame and a winding frame. For the potential we studied in section 4, the solutions are unphysical in both supergravity frame and winding frame. In section 5, we briefly considered the T-dual solutions in both frames and obtained the solutions of a constant usual diffeomorphic dilaton $\phi_0$ in an FRW universe in both a supergravity frame and in a winding frame respectively. In section 6, we found the basic conditions for three dark energy candidates in $D=4$ DFT cosmology. In section 7, we analyzed a more general potential $V(d) = A e^{nd}$, where $A$ can be arbitrary real numbers. More general results have been obtained. \\
A further line of study using string theory would be to use exceptional field theory (ExFT) [11] of which Double field theory (DFT) is a subset. Both geometrization of the  $p$-form fluxes in 10-/11$d$ supergravity and their unification with the metric of degrees of freedom would also be interesting lines for further research. DFT corresponds to the $NS-NS$ sector of 10$D$ supergravity, whilst ExFT would also include the $R-R$ sector of 10$D$ type $II$ supergravity or would work for 11$D$ supergravity. These theories are sometimes referred to as being “duality invariant”: in this sense, DFT captures T-duality and ExFT extends this to U-duality. There have been no studies so far of cosmology using exceptional field theory as proposed above and in future work we therefore intend to investigate the effects of ExFT cosmology.  

\appendix
\section{Appendix: A note concerning T-dualizing matter}
Matter is introduced in supergravity by the following action [3, 28]:
\begin{equation}\label{eq:A1}
S = \int d^D x \sqrt{-g} e^{-2\phi}f = \int dt \sqrt{-g_{tt}} F(\log a, \beta\sqrt{-g_{tt}}, \phi),
\end{equation} 
where $F$ is the free energy,
\begin{equation}\label{eq:A2}
F = \int d^{D-1} x a^{D-1} e^{-2\phi} f.
\end{equation}
We can consider a T-dual covariant generalization of $F$ by defining the following action:
\begin{equation}\label{eq:A3}
S = \int d^{D} x d^D {\tilde{x}} e^{-2\phi} \mathcal{F}, 
\end{equation}
with $\mathcal{F}$ dependent on both sets of coordinates. Then we can write
\begin{equation}\label{eq:A4}
S = \int dt \sqrt{-g_{tt}} \{\int d^{D-1} \tilde{x} d\tilde{t} F \} 
\end{equation}
or 
\begin{equation}\label{eq:A5}
S = \int d\tilde{t} \sqrt{-\frac{1}{g_{tt}}} \{\int d^{D-1} \tilde{x} dt F \},  
\end{equation}
where $F$ is also a function of both sets of coordinates. \\
Finally, the standard definitions of the energy and pressure of the system are derived as usual according to
\begin{equation}\label{eq:A6}
E(t, \tilde{t}) = -2 \frac{\delta F}{\delta g_{00}},  
\end{equation}
\begin{equation}\label{eq:A7}
P_i(t, \tilde{t}) = - \frac{\delta S}{\delta \ln a_i}.  
\end{equation}

All data generated or analysed during this study are included in this published article.


\end{document}